\documentclass[12pt]{article}
\usepackage{epsfig,amsfonts,amssymb}
\usepackage{hyperref}
\pdfoutput=1

\usepackage{cite}
\usepackage{cite}

\usepackage{comment}

\def\ZZZ{{\hbox{ Z\kern-1.6mm Z}}}
\def\RRR{{\hbox{ R\kern-2.4mm R}}}
\def\CCC{{\hbox{ C\kern-2.0mm C}}}
\def\zzz{{\hbox{z\kern-1mm z}}}

\newcommand{\nn}{\nonumber \\ }

\newcommand{\qeq}{{\hbox{=\kern-2.3mm ? \kern.5mm }}}
\renewcommand{\qeq}{=}

\newcommand{\LL}{{\cal L}}

\newcommand{\wt}{\widetilde}

\newcommand{\NN}{{\cal N}}

\newcommand{\be}{\begin{equation}}
\newcommand{\ee}{\end{equation}}
\newcommand{\ben}{\begin{eqnarray}\displaystyle}
\newcommand{\een}{\end{eqnarray}}

\newcommand{\refb}[1]{(\ref{#1})}
\newcommand{\p}{\partial}
\newcommand{\sectiono}[1]{\section{#1}\setcounter{equation}{0}}

\def\one{{\hbox{ 1\kern-.8mm l}}}
\def\zero{{\hbox{ 0\kern-1.5mm 0}}}

\newcommand{\bea}[1]{\begin{eqnarray}\label{#1} }
\newcommand{\eea}{\end{eqnarray}}

\newcommand{\eqref}{\refb}

\usepackage{bm}
\usepackage[table]{xcolor}

\begin{document}

\baselineskip 18pt

\thispagestyle{empty}
\begin{center}
    ~\vspace{5mm}
    
    {\Large \bf
    
 Supersymmetric Index for Small   Black Holes  
    }
    
    \vspace{0.4in}
    
    \centerline{\bf 
    Chandramouli Chowdhury,$^{1,2}$  Ashoke Sen,$^2$
    P Shanmugapriya,$^3$ 
    Amitabh Virmani$^{3}$
    }

    \vspace{0.4in}

    $^1$ Mathematical Sciences and STAG Research Centre, 
    University of Southampton, Highfield, Southampton SO17 1BJ, United Kingdom
     \vskip1ex
    $^2$ International Centre for Theoretical Sciences, 
    Bengaluru - 560089, India \vskip1ex
    $^3$ Chennai Mathematical Institute, 
    Kelambakkam, Tamil Nadu, India 603103    \vspace{0.1in}
    
    {\tt chandramouli.chowdhury@icts.res.in,
    ashoke.sen@icts.res.in, shanmugapriya@cmi.ac.in, avirmani@cmi.ac.in}
\end{center}

\vspace{0.4in}

\begin{abstract}

Supersymmetric elementary string states in the compactified heterotic string theory 
are described by small black holes that have
zero area event horizon. In this paper we compute the supersymmetric
index of such elementary string states 
using gravitational path integral. The dominant contribution to the
path integral comes from an Euclidean rotating black hole solution of the
supergravity theory with a finite area event
horizon, but  the logarithm of the index, computed from the saddle
point, vanishes. Nevertheless we show that the solution is singular on certain subspaces
of the horizon where higher derivative corrections can be important, and once the higher
derivative corrections are taken into account the solution could yield a finite result for
the logarithm of the index whose
form agrees with the microscopic results up to an overall numerical constant.
While the numerical constant is not
determined in our analysis,  we show that it is independent
of the details of the compactification and even 
the number of non-compact dimensions, in
agreement with the microscopic results.

\end{abstract}

\pagebreak

\baselineskip=18pt

\tableofcontents

\sectiono{Introduction}

The excitations of a string lead to
an infinite tower of massive states with large degeneracy. It has long been suspected
that sufficiently heavy states of the string can be identified as black holes and that the
logarithm of the
degeneracy of such string states may provide an explanation of the black hole 
entropy\cite{tHooft:1990fkf,Susskind:1993ws,Susskind:1994sm,Russo:1994ev,
Horowitz:1996nw,Chen:2021dsw}. 
However since the counting of states of the elementary string can be
carried out reliably at weak string coupling, while the description of the system as a 
black hole holds at large string coupling, quantitative comparison between
the microscopic and macroscopic calculations is difficult. 

This problem can be
overcome by considering a special class of elementary string states that preserve
some of the supersymmetries of the theory and hence saturate the
BPS bound\cite{Dabholkar:1989jt}. 
For such states there are general arguments that the log of the
degeneracy computed at weak coupling remains the same at strong coupling and hence 
can be compared with the entropy of the black hole carrying the same charges. One finds
however that the corresponding black hole has vanishing area of the event horizon and hence
the entropy vanishes. 
It was argued in \cite{Sen:1995in,Peet:1995pe,Sen:1997is} 
that once we take into account the higher derivative corrections
to the effective action, the black hole may acquire a finite entropy. Using the symmetries
of the theory and a scaling argument, one could determine the dependence of the entropy
on the charges  up to an overall numerical factor whose value depends on the details
of the higher derivative corrections\cite{Sen:1995in}. 
The dependence of the entropy on the charges,
obtained this way, precisely matches with the known result from the microscopic 
counting\cite{Sen:1995in,Peet:1995pe,Sen:1997is}. The numerical constant was
determined later in \cite{Dabholkar:2004yr} under the assumption that only
certain four derivative terms 
contribute to the entropy of the black hole.

There is however one caveat in this analysis. The analysis described above relies on
the degeneracy of BPS states being protected under quantum corrections so that it takes the
same value for weak and strong coupling. However it is not the degeneracy, but a closely
related quantity called the supersymmetric index, that is protected this way. So we need to
make an extra assumption, namely that the degeneracy and the
index should be equal both at
strong and weak coupling. In the weak coupling limit, where the states can be 
described as excitations of the fundamental string, 
this is known to be the case for compactification of heterotic string theory. However for type II
string compactification this fails -- the index vanishes while the degeneracy continues to be
large. At strong coupling where the black hole description is valid, there is no general argument
showing the equality of degeneracy and the index either in the heterotic string theory or in
type II string theory.  See {\it e.g.} section 5 of
\cite{Sen:2005ch} for some discussion on these issues.

The goal of this paper is to circumvent this problem by directly analyzing the index of
fundamental string states in heterotic string theory compactified on $(10-D)$
dimensional torus and other manifolds. On the microscopic side the computation is
straightforward and the result has been known for many years. On the macroscopic
side we use a procedure for computing index developed recently in \cite{2107.09062}, 
following
earlier work on $AdS$ spaces\cite{Cabo-Bizet:2018ehj}. 
This procedure has been successful in reproducing
correctly the leading result\cite{2107.09062,Boruch:2023gfn}
as well as logarithmic correction to the logarithm of the
index of black holes with regular finite area event 
horizon\cite{H:2023qko,Anupam:2023yns}. We use this method to calculate
the leading contribution to the logarithm of the index of small black holes.

It turns out that the saddle point that contributes to the log of the index of small black holes
is not a small black hole with zero area event horizon
but a rotating Euclidean black hole with finite area  event
horizon. Nevetheless, when one follows the algorithm for computing the index using this
black hole saddle point, one finds that in the two derivative theory the index vanishes, just
as the entropy of the small black holes vanished in the two derivative theory. One finds
however that despite having an event horizon of finite area,
this black hole solution is not completely smooth, -- it has singularity on a
subspace of the horizon where higher derivative corrections may become important. We
carefully analyze the geometry near the singularity, 
and by using the symmetries of the theory and a scaling
argument very similar to the ones used in \cite{Sen:1995in}, 
we can completely determine the  
corrected logarithm of the index  up to an overall
numerical factor.  This dependence of the index 
on the charges found this way turns out to be in perfect agreement with the
microscopic results for the index. We furthermore show that the undetermined 
numerical factor 
is independent of the dimension $D$ of the non-compact part of the space-time
even though the full black hole
solution depends non-trivially on $D$. This is also in
agreement with the microscopic results.

The rest of the paper is organized as follows. In section \ref{smicro} we review the
microscopic results for the index of fundamental string states. In section \ref{sstrategy} we
describe the general strategy that we shall use for the macroscopic computation of the log
of the index. In section \ref{smacro} we describe the rotating black hole solution that
contributes to the index and show that the log of the index computed from this solution
vanishes in the two derivative theory. We also show that the
solution is singular on a subspace of the
horizon. In section \ref{ssingular} we zoom in on the region near the singularity and 
express the solution in a rescaled coordinate system that is useful for
studying the solution near the singularities. In section \ref{swald} we
present the argument based on scaling and the symmetries that determines the
higher derivative correction to the log of the index up to an overall numerical
constant. We show that the dependence of this correction on the charges carried by the
state is precisely what is expected from the microscopic analysis. In section \ref{sgen} we
generalize these results to other compactifications of heterotic string theory.

Since the paper is technical, for the benefit of the reader we would like to point to
the final results of our paper that
are contained in section \ref{swald}. Eqs.\refb{eF1}-\refb{eF6}
describe the solution near the singularity. These 
configurations are universal, i.e.\ independent
of the parameters of the solution like the charges or the asymptotic values of the moduli,
except for two factors. First the string metric \refb{eF1}
contains a direct sum of a universal four dimensional metric and a nearly flat $(D-4)$
dimensional metric $b^2d\Omega^{D-4}$, 
whose only role in our analysis is to provide a factor of the volume
of the $(D-4)$ dimensional 
space, proportional to $b^{D-4}$,
in the evaluation of the action. The second source of non-universality
is an additive constant $-\ln (g_s^{-2} m_0 b^{D-4})$ 
in the expression for the dilaton field $\Phi$ as given in \refb{eF6},
which also provides an overall multiplicative factor in the action but does not
affect our analysis in any other way. Expressing these non-universal factors in terms
of the charges carried by the black hole, we arrive at the result \refb{esmac2} which is our final
result for the logarithm of the index. Here $K$ is an unknown numerical factor that can
in principle be determined with a better understanding of the higher derivative terms in
the action, or, equivalently, a better understanding of the world-sheet $\sigma$ model
with target space determined by the solutions \refb{eF1}-\refb{eF6}.

Ref.~\cite{Chen:2021dsw} gives an argument that 
the black hole solution describing elementary string
states and the string star solution of \cite{Horowitz:1996nw} 
belong to the same moduli space parametrized
by the mass of the black hole. If a similar relation can be established for 
charged rotating black
holes, then we may be able to use the correspondence to map the problem of computing
$K$ to the problem of computing the entropy of a rotating string star solution. It will be
interesting to explore this avenue.

\sectiono{Microscopic results for small black hole entropy} \label{smicro}

We consider heterotic string theory compactified on a $(10-D)$ dimensional torus so that we
have a theory in $D$ non-compact space-time directions.  The vacuum of this
theory is characterized by the string coupling $g_s$ that can be identified as the 
vacuum expectation value of $e^{\Phi/2}$, $\Phi$ being the dilaton field
and the vacuum expectation value of a $(36-2D)\times (36-2D)$ 
matrix valued scalar field $M$, satisfying 
\be \label{edefML}
MLM^T = L, \qquad M^T=M, \qquad L = \pmatrix{-I_{26-D} & 0 \cr 0 & I_{10-D}}\, ,
\ee
where $I_n$ denotes $n\times n$ identity matrix. We shall denote by $\Phi_\infty$ and
$M_\infty$ the asymptotic values of the fields $\Phi$ and $M$.

The elementary string states in
this theory carry $36-2D$ different charges, which we can divide into a $(10-D)$ dimensional
charge vector $Q_R$ and a $(26-D)$ dimensional charge vector $Q_L$. 
If we denote by $Q$ the full $(36-D)$
dimensional charge vector, then we can write
\be 
Q_R = {1\over 2} (M_\infty+L) Q, \qquad Q_L = {1\over 2} (M_\infty-L) Q\, .
\ee
We also define,
\be\label{edefqr2}
Q_R^2 \equiv  {1\over 2} Q^T (M_\infty+L) Q, \qquad 
Q_L^2\equiv {1\over 2} Q^T (M_\infty-L) Q, \qquad Q^2 \equiv Q^T L Q = Q_R^2-Q_L^2\, .
\ee
Among the elementary string states there
are a special class of states, known as half
BPS states, that are invariant under eight of the sixteen
unbroken supersymmetries of the theory. The mass of such a state, measured in the string
metric, is given by 
\be\label{ebps}
m_{BPS} = \sqrt{Q^T (M_\infty+L) Q} = \sqrt{2\, Q_R^2}\, .
\ee
At this point we need to describe our convention. We shall work in the $\alpha'=1$
unit so that if we have a fundamental string carrying $n$ units of momentum and $w$
units of winding along an internal circle of radius $R$, it has mass $wR + n/R$.  In
our convention $Q_R=(wR + n/R)/\sqrt 2$ and $Q_L = (wR - n/R)/\sqrt 2$ for this state so that
\refb{ebps} holds 
and we have
$Q_R^2-Q_L^2 = 2nw$.

The microscopic entropy of such BPS states, defined as the log of the degeneracy of
the states, 
can be computed and yields the result (see {\it e.g.} \cite{Russo:1994ev,Sen:1995in})
\be\label{emicro}
S_{micro} =4\pi \sqrt{Q_R^2-Q_L^2\over 2 } =2\sqrt 2 \pi \sqrt{Q^2}\, ,
\ee
in the limit of large charges.
However this is not protected under quantum corrections. The more relevant quantity 
for our analysis is
an appropriate index, known as the fourth helicity trace index, defined as
\be\label{ehelicity}
  {1\over 4!} \, Tr_Q \left[ e^{-\beta (H - m_{BPS})} (-1)^F (2 J)^4\right]\, ,
\ee
where $Tr_Q$ denotes trace over all states carrying charge $Q$ and zero momentum,
$(-1)^F$ takes value 1 and $-1$ for bosonic and fermionic states respectively, $H$
is the Hamiltonian and
$J$ is the component of the angular momentum carried by the state in some fixed
two dimensional plane. $\beta$ is an arbitrary parameter labelling the 
inverse temperature of the
system but the index is supposed to be independent of $\beta$ since  only those states
whose $H$ eigenvalue is equal to $m_{BPS}$
contribute  to the trace.
For heterotic string theory on $T^{10-D}$ this quantity can be
shown to be equal to the degeneracy in the weak coupling limit, and so the log of the
index is given by \refb{emicro}. However this index can also be shown to be protected
under quantum corrections and hence takes the same value at strong coupling.

Our goal is to try to reproduce \refb{emicro} 
from macroscopic analysis, i.e.\ from the
analysis of the classical geometry describing a solution carrying these charges.

\sectiono{Strategy for macroscopic computation of small black hole entropy}
\label{sstrategy}

In any two derivative theory in $D$ space-time dimensions containing the metric
$g_{\mu\nu}$, 2-form field $B_{\mu\nu}$, $U(1)$ gauge fields $\{A_\mu^{(i)}\}$
and scalar fields $\{\phi_\alpha\}$, there is a scaling relation under which
\be\label{e1}
g_{\mu\nu}\to \lambda^2\, g_{\mu\nu}, \quad 
B_{\mu\nu}\to \lambda^2\, B_{\mu\nu}, \quad
A^{(i)}_{\mu}\to \lambda
\, A^{(i)}_{\mu}, \quad \phi_\alpha
\to \phi_\alpha,
\ee
and the action scales as $\lambda^{D-2}$. For a classical black hole
solution carrying mass $m$, charges $Q$ and angular momenta $J$, 
if we scale the fields as in \refb{e1},  
the various quantities parametrizing the solution scale as
\be\label{e22}
m\to \lambda^{D-3}\, m, \quad Q\to \lambda^{D-3}\, Q, \quad
J\to \lambda^{D-2}\, J\, .
\ee
Under this scaling the Bekenstein-Hawking 
entropy $S_{BH}$ of a regular black hole scales in the same way as the
action:
\be\label{e23}
S_{BH}\to  \lambda^{D-2} \, S_{BH}\,  .
\ee
These results can be derived as follows. First note that in the usual coordinate 
system where the asymptotic metric is taken to be $\eta_{\mu\nu}$, the metric does
not scale as given in \refb{e1}. For this we introduce new coordinate system $\tilde x^\mu
= x^\mu/\lambda$ so that the the asymptotic metric has the form $\lambda^2 \eta_{\mu\nu}
d\tilde x^\mu  d\tilde x^\nu$ in accordance with the scaling laws in \refb{e1}. In this
coordinate system the usual fall-off properties take the form:
\ben
&& (1+g_{tt}) \propto {m\over r^{D-3}} \ \Rightarrow\ (1+  \lambda^{-2} g_{\tilde t
\tilde t}) \propto {m \over \lambda^{D-3}  \tilde r^{D-3}},
\qquad 
g_{t x^i} \propto {J\over r^{D-2}}\ \Rightarrow\  \lambda^{-2}
g_{\tilde t \tilde x^i} \propto {J \over \lambda^{D-2}  \tilde r^{D-2}},
\nonumber \\
&& A_{t} \propto {Q\over r^{D-3}}\ \Rightarrow\  
\lambda^{-1} A_{\tilde t} \propto {Q\over \lambda^{D-3} \tilde r^{D-3}}\, .
\een
We now see that the $\lambda$ dependence of
these equations can be compensated by
scaling the fields as in \refb{e1} and $m,Q,J$ as in \refb{e22}. 

We can take
the macroscopic limit by taking $\lambda$ to be large.
Note that under the same scaling \refb{emicro} scales by a factor of
$\lambda^{D-3}$, leading to an
apparent contradiction. This issue was resolved in \cite{Sen:1995in}
by noting that the classical
BPS black hole in the two derivative supergravity theory has zero area event horizon 
and hence vanishing entropy and we need to take into account higher derivative corrections
to get the correct scaling of the entropy. We shall carry out a similar analysis in this paper,
but focussing from the beginning on the computation of the index on the macroscopic side
as well.

We now outline the main  steps in our analysis.
\begin{enumerate}
\item Since we are interested in computing the index, we follow the strategy described
in \cite{2107.09062}, where we begin with an electrically charged 
black hole carrying a charge vector $Q$ and angular momentum $J$ in a two dimensional
plane. The angular momentum $J$ is adjusted so that the chemical potential $\Omega$
dual to this angular momentum, also called the angular velocity, is given by 
\be\label{ebetaomega}
\Omega = -2\pi \mathrm{i} /\beta\, ,
\ee
where $\beta$ is the inverse temperature of the black hole.  The effect of this chemical
potential is to insert a factor of $e^{2\pi \mathrm{i} J}$ into the path integral, which in turn plays
the role of $(-1)^F$ that appears in \refb{ehelicity}. Therefore this rotating black hole
solution provides the dominant saddle point in the computation of the index.
Then the leading semiclassical result $S_{macro}$ for the
log of the index, computed from the gravitational
path integral, is given
by 
\be \label{efull}
S_{macro} = S_{wald}+2\pi \mathrm{i} J\, ,
\ee
where $S_{wald}$ is the Wald entropy of the black hole, which can also be computed
following the approach of Gibbons and Hawking where we first compute the partition
function by computing the path integral over all the fields and then use the
appropriate thermodynamic relations to extract the entropy\cite{Wald:1993nt}.  
We shall denote by
$S_{wald}$ the expression for the black hole entropy including the effect of
higher derivative corrections, irrespective of which way it is calculated.
For two derivative theories this
agrees with the Bekenstein-Hawking entropy $S_{BH}$
of the black hole given by $A/(4G_N)$ where $A$
is the horizon area and $G_N$ is the Newton's gravitational constant.
The $2\pi \mathrm{i} J$ term in \refb{efull} is
the effect of the chemical potential that inserts the explicit factor of $e^{2\pi\mathrm{i} J}$
in the path integral. Note that according to \refb{ehelicity} we also need to insert
a factor of $(2J)^4$ into the path integral, but as discussed in \cite{H:2023qko,Anupam:2023yns},
this factor just soaks up the fermion zero modes associated with broken supersymmetries
in the path integral and does not have any other effect.
\item We then show that for this solution
\be\label{e2}
S_{BH} + 2\pi \mathrm{i} J =0\, .
\ee
If this had been non-zero, then according to \refb{e22}, \refb{e23}
this would scale as $\lambda^{D-2}$. On the other hand \refb{emicro} and
\refb{e22} shows that $S_{micro}$ scales as $\lambda^{D-3}$. This would lead
to an obvious contradiction which is avoided by \refb{e2}. 

\item 
We can now ask whether higher derivative corrections to \refb{efull} could resolve the
disagreement. Since $J$ is measured from the fields at infinity where the higher derivative
corrections can be ignored, we focus on $S_{wald}$.
If the horizon had been smooth, then the four derivative corrections to $S_{wald}$
would be of order $\lambda^{D-4}$ 
according to \refb{e1} since two extra derivatives need to be
contracted with a $g^{\mu\nu}$. 
This would be inconsistent with \refb{emicro} which scales
as $\lambda^{D-3}$.
We show that the solution is singular near a subspace, and that in the neighbourhood of
this subspace 
we can choose a coordinate system in which the string frame 
metric becomes independent of the
charges and the dilaton takes the form
\be\label{e3}
e^{-\Phi} = \sqrt{Q_R^2-Q_L^2} \ F
\ee
where $F$ is a function of the coordinates that does not depend on the charges.
Since $e^{\Phi/2}$ expectation value gives the string coupling, \refb{e3} shows
that for large charges, the string coupling at the horizon is small. 
Hence we can restrict our
analysis to tree level string theory. On the other hand since the metric and $\p_\mu\Phi$ are of
order unity (having no dependence on the charges) we see that the $\alpha'$ corrections
are of order unity, and we need to take into account all order corrections in $\alpha'$ to
compute the entropy. However these corrections are completely universal, independent
of the charges.
\item This argument shows that if the higher derivative corrections generate a finite entropy,
then the dependence of the entropy on the charges comes solely from the 
$\sqrt{Q_R^2-Q_L^2}$ factor in \refb{e3}. This will appear as an overall multiplicative factor
in the expression for the on-shell action and hence the entropy. So we can write
\be\label{emacro}
S_{macro} = C\, \sqrt{Q_R^2-Q_L^2}\, ,
\ee 
in agreement with \refb{emicro}. Here $C$ is an unknown numerical constant that depends on
the details of how the $\alpha'$ corrections modify the near horizon geometry.
However $C$ is independent of $\Phi_\infty$ and $M_\infty$. We also show that $C$ is
independent of $D$, in agreement with \refb{emicro}.
\end{enumerate}

Finally we would like to remark that since the macroscopic  computation of the
index will be done using the full space-time
geometry, it implicitly uses the grand canonical ensemble, i.e.\ instead
of computing the trace over fixed charge states as in \refb{ehelicity} we sum over all charged
states weighted by $e^{-\beta \mu.Q}$ and then perform the Legendre transform
of the logarithm of
the result with respect to the chemical potential $\mu$. Of course we shall not see
this being done explicitly, but the gravitational formula \refb{efull}
for the index that we shall use
does this automatically.  If the index is a smooth
function of the charges then the result of this computation differs from the actual index
by subleading logarithmic terms that we do not keep track of. However if the index is
not a smooth function of the charges, {\it e.g.} if it switches sign as we move from one
element of the charge lattice to its neighbouring element, then the procedure described
above can give results that
differ significantly from the actual index, -- see {\it e.g.} \cite{Cabo-Bizet:2018ehj}
for similar issues for AdS back holes. For heterotic string compactification we know that
the microscopic index is a smooth function of the charges for large charge.
We shall assume that this holds on the macroscopic side as well.

\renewcommand{\bea}{\ben}
\renewcommand{\eea}{\een}

\sectiono{The black hole solution} \label{smacro}

We begin by writing down the two derivative action describing the massless
bosonic fields in heterotic string theory compactified on 
$T^{10-D}$~\cite{Maharana:1992my}:
\bea \label{action}
S &=& C_D\int d^D x \sqrt{-\det G} \, e^{-\Phi} \, \Big[ R_G + G^{\mu\nu}
\p_\mu \Phi \p_\nu\Phi +{1\over 8} G^{\mu\nu} \, \mathrm{Tr}(\p_\mu M L\p_\nu ML)
\nonumber \\
&& -{1\over 12} G^{\mu\mu'} G^{\nu\nu'} G^{\rho\rho'} H_{\mu\nu\rho}
H_{\mu'\nu'\rho'} - G^{\mu\mu'} G^{\nu\nu'} F^{(j)}_{\mu\nu} \, (LML)_{jk}
\, F^{(k)}_{\mu'\nu'} \Big] \, ,
\eea
where,
\be 
F^{(j)}_{\mu\nu} = \p_\mu A^{(j)}_\nu - \p_\nu A^{(j)}_\mu \, ,
\ee
\be \label{eH}
H_{\mu\nu\rho} = 
\p_\mu B_{\nu\rho} + 2 A_\mu^{(j)} L_{jk} F^{(k)}_{\nu\rho}
+\p_\nu B_{\rho\mu} + 2 A_\nu^{(j)} L_{jk} F^{(k)}_{\rho\mu}
+\p_\rho B_{\mu\nu} + 2 A_\rho^{(j)} L_{jk} F^{(k)}_{\mu\nu}, \label{def_H}
\ee
and $R_G$ is the Ricci scalar associated with the metric
$G_{\mu\nu}$. $M$ and $L$ have been introduced in \refb{edefML}.
$C_D$ is an arbitrary $D$-dependent 
constant, which can be absorbed into the dilaton field $\Phi$. 
This action is invariant under an $O(26-D, 10-D)$ symmetry under 
which
\be\label{eodd}
M\to \Omega M \Omega^T, \qquad  A_\mu^{(j)} \to \Omega_{jk} A_\mu^{(k)}, \qquad
\forall \ \Omega \ \hbox{such that} \ \Omega^T L \, \Omega=L\, .
\ee
Furthermore this symmetry is known to survive $\alpha'$ correction to the effective action,
i.e.\ this is an exact symmetry of tree level string theory\cite{Sen:1991zi,Hassan:1991mq}.

$G_{\mu\nu}$ appearing in \refb{action} is the string frame metric. The 
canonical Einstein frame metric
$g_{\mu\nu}$ is related to this via the equation:
\be\label{edefcan}
g_{(can)\mu\nu} =g_s^{4/(D-2)} e^{-2\Phi/(D-2)} G_{\mu\nu}\, ,
\ee
where $g_s^2$ is the asymptotic value of $e^\Phi$. The $g_s$ dependent factor is 
normally not included in the definition of the Einstein frame metric, but we have 
included it here so that if $G_{\mu\nu}$ approaches $\eta_{\mu\nu}$ asymptotically,
so does $g_{\mu\nu}$.
We shall always work with the string frame metric unless mentioned otherwise.

\subsection{The solution}
Let $\alpha$ and $\gamma$ be two boost parameters.  Let $\vec n$ be a $(26-D)$
dimensional unit vector and $\vec p$ be a $(10-D)$ dimensional unit
vector. The general non-extremal solution in $D$-dimensions with single rotation takes the form~\cite{Horowitz:1995tm}:
\bea \label{metric_non_extremal}
ds^2 &\equiv& G_{\mu\nu} dx^\mu dx^\nu \nonumber \\
&=&  (\rho^2+a^2\cos^2\theta)\Big\{ - \Delta^{-1}
(\rho^2+a^2\cos^2\theta -2m\rho^{5-D}) dt^2 +
(\rho^2+a^2-2m\rho^{5-D})^{-1} d\rho^2
+d\theta^2 \nonumber \\
&& +\Delta^{-1} \sin^2\theta [\Delta + a^2\sin^2\theta (\rho^2 + a^2\cos^2
\theta + 2m\rho^{5-D} \cosh\alpha\cosh\gamma)] \, d\phi^2
\nonumber \\ &&
- 2\Delta^{-1} m\rho^{5-D} a \sin^2\theta (\cosh\alpha + \cosh\gamma) dt
d\phi + \rho^2 \cos^2\theta
(\rho^2+a^2\cos^2\theta)^{-1} d \Omega^{D-4} \Big\}\, ,
\eea
where,
\be \label{Delta}
\Delta = (\rho^2 + a^2\cos^2\theta)^2 + 2m \rho^{5-D}
(\rho^2 + a^2 \cos^2\theta)
(\cosh\alpha \cosh\gamma -1) + m^2\rho^{10-2D} (\cosh\alpha -\cosh\gamma)^2.
\ee
$\{t, \rho, \theta, \phi\}$ are the four spacetime coordinates and $d \Omega^{D-4}$ is the round metric on a $(D-4)$- dimensional unit sphere. 
The dilaton supporting the solution is:
\be \label{dilaton}
\Phi=2\ln g_s + {1\over 2} \ln {(\rho^2+a^2\cos^2\theta)^2\over \Delta}.
\ee
The U(1) vector fields $A^{(j)}$ for $1\le j\le (26-D)$ are determined by the unit vector $\vec n$:
\bea \label{vector}
A^{(j)}_t &=& -{n^{(j)} \over \sqrt 2} \Delta^{-1} m\rho^{5-D}
\sinh\alpha \{ (\rho^2+
a^2\cos^2\theta)\cosh\gamma + m\rho^{5-D} (\cosh\alpha - \cosh\gamma)\}, \nonumber \\
A^{(j)}_\phi &=& {n^{(j)}\over \sqrt 2} \Delta^{-1} m \rho^{5-D} a \sinh\alpha
\sin^2\theta \{ \rho^2 + a^2\cos^2\theta + m\rho^{5-D}\cosh\gamma (\cosh
\alpha -\cosh\gamma)\}\, , \nonumber \\
\eea
and the remaining vector fields $A^{(j)}$ for $j\ge (27-D)$ are determined by the unit vector $\vec p$:
\bea
A^{(j)}_t &=& -{p^{(j-26+D)}\over \sqrt 2} \Delta^{-1} m\rho^{5-D}
\sinh\gamma \{ (\rho^2+
a^2\cos^2\theta)\cosh\alpha + m\rho^{5-D}
(\cosh\gamma - \cosh\alpha)\},  \nonumber \\
A^{(j)}_\phi  &=& {p^{(j-26+D)}\over \sqrt 2} \Delta^{-1} m \rho^{5-D} a \sinh\gamma
\sin^2\theta \{ \rho^2 + a^2\cos^2\theta + m\rho^{5-D}\cosh\alpha (\cosh
\gamma -\cosh\alpha)\}. \nonumber \\
\eea
The two-form field takes the form,
\be \label{two-form}
B_{t\phi} = \Delta^{-1} m\rho^{5-D} a
\sin^2\theta (\cosh\alpha -\cosh\gamma)
\{ \rho^2 + a^2\cos^2\theta + m\rho^{5-D}(\cosh\alpha \cosh\gamma -1)\} \,,
\nonumber \\
\ee
and the matrix of scalars takes the form,
\be \label{matrix_M}
M = I_{36-2D} +
\pmatrix{
 P n n^T & U n p^T \cr  U p n^T  & P pp^T}\, ,
\ee
where,
\bea
P &=& 2\Delta^{-1} m^2\rho^{10-2D} \sinh^2\alpha \sinh^2\gamma \, ,
\nonumber \\
U &=& - 2 \Delta^{-1} m\rho^{5-D} \sinh\alpha \sinh\gamma \{ \rho^2 + a^2
\cos^2\theta + m\rho^{5-D} (\cosh\alpha \cosh\gamma -1) \} \, .
\eea
Since $P$ and $U$ fall off for large $\rho$, we see from \refb{matrix_M} that $M_\infty
=I_{36-2D}$. However using the symmetry transformation \refb{eodd} we can generate any
other $M_\infty$ from this solution, and hence setting $M_\infty$ to $I_{36-2D}$ does not
restrict the solution in any way.

\subsection{BPS limit and Euclidean continuation} \label{sbps}
The BPS limit is obtained by taking $m \to 0$, $\gamma \to \infty$ keeping 
$ m_0 \equiv m \cosh \gamma$ finite. The BPS solution is also written in 
\cite{Horowitz:1995tm}. 
Let us in addition consider the  analytic continuation to Euclidean space, together with
a redefinition of the rotation parameter $a$ and the fields:
\ben \label{eanalytic} 
&t = \mathrm{i} \, \tau, &  \nonumber \\  
&a  =  \mathrm{i} \, b, \qquad J = \mathrm{i} J_E,  \qquad \Omega = -
\mathrm{i} \, \Omega_E, &  \nonumber\\
& A^{(j)} = \mathrm{i}  A_{(E)}^{(j)},& \nonumber \\
& B = - B_{(E)}.
\een 
The last analytic continuation is required for consistency with \refb{def_H}.
This gives the following Euclidean solution,
\bea \label{emetric}
ds^2 &=& (\rho^2-b^2\cos^2\theta) \Big\{ \Delta^{-1}
(\rho^2-b^2\cos^2\theta) d\tau^2 +
(\rho^2-b^2)^{-1} d\rho^2
+d\theta^2 \nonumber \\
&& +\Delta^{-1} \sin^2\theta [\Delta -b^2\sin^2\theta (\rho^2 -b^2\cos^2
\theta +2m_0\rho^{5-D}\cosh\alpha)] \, d\phi^2
\nonumber \\  &&
+ 2\Delta^{-1} m_0\rho^{5-D} b \sin^2\theta d\tau
d\phi + \rho^2\cos^2\theta(\rho^2 -
b^2\cos^2\theta)^{-1}d\Omega^{D-4}\Big\},
\eea
where now
\be 
\Delta = (\rho^2 - b^2\cos^2\theta)^2 + 2m_0 \rho^{5-D}\cosh\alpha \,
(\rho^2 - b^2 \cos^2\theta)
+ m_0^2\rho^{10-2D} \, ,
\ee
\be \label{e320}
\Phi=2\ln g_s+ {1\over 2} \ln {(\rho^2- b^2\cos^2\theta)^2\over \Delta}.
\ee
The gauge fields are given by
\bea \label{e323} 
A_{(E)}^{(j)}{}_\tau &=& -{n^{(j)}\over \sqrt 2} \Delta^{-1} m_0\rho^{5-D}
\sinh\alpha (\rho^2 - b^2\cos^2\theta)\, , \nonumber \\
A_{(E)}^{(j)}{}_\phi &=& -{n^{(j)}\over \sqrt 2} \Delta^{-1} (m_0)^2 \rho^{10-2D}
b \sinh\alpha \sin^2\theta , \qquad \hbox{for $1\le j\le 26-D$}\, ,
\eea
and 
\bea \label{e325}
A_{(E)}^{(j)}{}_\tau &=& -{p^{(j-26+D)}\over \sqrt 2} \Delta^{-1} m_0\rho^{5-D} \{ (\rho^2-
b^2\cos^2\theta)\cosh\alpha + m_0\rho^{5-D} \} \, , \nonumber \\
A_{(E)}^{(j)}{}_\phi &=& 
{p^{(j-26+D)}\over \sqrt 2} \Delta^{-1} m_0 \rho^{5-D} b
\sin^2\theta \{ \rho^2 - b^2\cos^2\theta + m_0\rho^{5-D}\cosh\alpha\} , \nonumber \\
&& \hskip 1in  \hbox{for $ j\ge 27-D$}\, .
\eea
The two form field takes the form,
\be
B_{(E)\tau \phi} = - \Delta^{-1} m_0\rho^{5-D} b
\sin^2\theta
\{ \rho^2 - b^2\cos^2\theta + m_0\rho^{5-D}\cosh\alpha \} \, ,
\ee
and the scalar fields are given by,
\be \label{eM}
M = I_{36-2D} +
\pmatrix{
 P nn^T & U n p^T \cr U p n^T & P pp^T} \, ,
\ee
where
\bea  \label{ePQ}
P &=& 2\Delta^{-1} m_0^2\rho^{10-2D} \sinh^2\alpha,  \nonumber \\
U &=& - 2 \Delta^{-1} m_0\rho^{5-D} \sinh\alpha \{ \rho^2 - b^2
\cos^2\theta + m_0\rho^{5-D} \cosh\alpha \}.
\eea

Note that with the redefinitions given in \refb{eanalytic} the solution is real. The horizon
is at $\rho=b$ and the relevant region of space-time 
is $\rho\ge b$. Also the solution is singular at
$\rho=b$, $\cos^2\theta=1$. This is easiest to see by examining the dilaton given in
\refb{e320}, which blows up at these values.

We now compute the electric charges carried by the solution. For this
we note that for large $\rho$, we have
\be
F^{(j)}_{(E)\rho\tau} \simeq \p_\rho A^{(j)}_{(E)\tau} = \cases{ 
(D-3) {n^{(j)}\over \sqrt 2} m_0 \, \rho^{2-D}\, 
\sinh\alpha \quad \hbox{for $1\le j\le 26-D$}\cr 
(D-3) {p^{(j-26+D)}\over \sqrt 2} m_0\, \rho^{2-D}\, 
\cosh\alpha \quad \hbox{for $27-D\le j\le 36-D$}
}\, .
\ee
We define the electric charges $Q^{(j)}$ as the integral of ${1\over 2}{\delta S\over \delta F^{(j)}_{(E)\rho\tau}}$
over the $(D-2)$ dimensional sphere at a large but
fixed value of $\rho$.\footnote{We have used the factor of 2 in the denominator to compensate
for the fact that the gauge field kinetic term in \refb{action} does not have the usual $1/4$.
We shall see that this normalization is consistent with the convention of section \ref{smicro}.} 
Using the fact that $e^{-\Phi_\infty}
=1/g_s^2$ and $M_\infty = I_{36-2D}$, 
\be
Q = \sqrt 2 \, (D-3)\, C_D \, g_s^{-2}\, m_0\, \omega_{D-2} \, \pmatrix{ n\sinh\alpha\cr p\cosh\alpha}
\equiv \pmatrix{Q_L\cr Q_R}\, ,
\ee 
where $\omega_k$ represents the volume of a unit sphere in $k$ dimensions.
Since $n$ and $p$ are unit vectors, we get
\be\label{echarges}
\sqrt{Q_R^2 - Q_L^2} = \sqrt{Q^T L Q} = 
\sqrt 2 \, (D-3)\, C_D \, g_s^{-2}\, m_0\, \omega_{D-2} \, .
\ee

We can also calculate the ADM mass of the black hole. For this we use \refb{emetric},
\refb{e320} and \refb{edefcan} to write down the expression for the canonical Einstein metric:
\bea \label{eforadm}
ds_{(can)}^2 &=& (\rho^2-b^2\cos^2\theta) \left({\rho^2-b^2\cos^2\theta\over \Delta^{1/2}} 
\right)^{-2/(D-2)} \nonumber \\ &&
\Big\{ \Delta^{-1}
(\rho^2-b^2\cos^2\theta) d\tau^2 +
(\rho^2-b^2)^{-1} d\rho^2
+d\theta^2 \nonumber \\
&& +\Delta^{-1} \sin^2\theta [\Delta -b^2\sin^2\theta (\rho^2 -b^2\cos^2
\theta +2m_0\rho^{5-D}\cosh\alpha)] \, d\phi^2
\nonumber \\  &&
+ 2\Delta^{-1} m_0\rho^{5-D} b \sin^2\theta d\tau
d\phi + \rho^2\cos^2\theta(\rho^2 -
b^2\cos^2\theta)^{-1}d\Omega^{D-4}\Big\},
\eea
Using this and eq.(14) of \cite{Emparan:2008eg},  we get
\be
m_{ADM} =  {(D-2) \, \omega_{D-2}\over 16\pi G_N}\lim_{\rho\to\infty} \rho^{D-3} (1-g_{\tau\tau})=
{\omega_{D-2}\over 8\pi G_N} \, (D-3)\, m_0\cosh\alpha \, ,
\ee
where the Newton's constant $G_N$ for the action \refb{action} is given by
\be
G_N = {1\over 16\pi C_D \, g_s^{-2}}\, .
\ee
This gives
\be\label{eadm2}
m_{ADM} = 2\, \omega_{D-2} \, C_D \, g_s^{-2}\, (D-3)\, m_0\cosh\alpha = \sqrt{2\, Q_R^2}\, ,
\ee
in agreement with \refb{ebps}.

The overall normalization of the charges depends on the normalization of the U(1)
gauge fields appearing in \refb{action}. We have chosen the normalization so that we get
\refb{eadm2}.
We have also carefully kept track
of all dependence on the asymptotic moduli and the dimension $D$. In particular, if we use
the transformation \refb{eodd} to generate a solution with some other
$M_\infty$, the dependence
of $m_{ADM}$ on $M_\infty$ and the charge $Q$ will be captured by the dependence of
$Q_R^2$ on $M_\infty$ and $Q$ via \refb{edefqr2},
since \refb{edefqr2} is invariant under
$O(26-D,10-D)$ transformation.

The area of the $\rho =b$ surface in the Einstein frame is
\be
A_H = m_0 b \,\omega_{D-2} \, .
\ee
The Bekenstein-Hawking entropy of the black hole is given by
\be
S_{BH} = \frac{A_H}{4G_N}={m_0 b\, \omega_{D-2} \over 4G_N}\, .
\ee
The Euclidean
angular momentum $J_E$ of the black hole is,
\be
J_E = \frac{\omega_{D-2}}{8 \pi G_N} m_0 b.
\ee
Clearly, 
\be\label{e2new}
S_{BH} - 2 \pi J_E = 0. 
\ee
This reproduces \refb{e2} with the identification given in \refb{eanalytic}.

\subsection{Periodicity of the $\tau$ circle}

To understand the periodicity of the $\tau$ circle, let us introduce
\ben\label{e443}
\widetilde \rho &= \sqrt{\rho-b}, \nonumber \\
\phi &= \widetilde \phi - \frac{b^{D-4} }{m_0}\tau.
\een
With these coordinate changes, the relevant part of the metric  near $\widetilde \rho = 0$
is  
\bea
ds^2 &\simeq&  2   b \sin^2 \theta \left\{ \left( d\widetilde \rho^2 + \widetilde \rho^2 \frac{b^{2D-8}}{m_0^2} d \tau^2\right) \right. \nn \\  
& & 
  \left. + \frac{b}{2} \left( d\theta^2 + \frac{m_0^2}{m_0^2 + 2 b^{D-3} m_0 \cosh \alpha \sin^2 \theta  + b^{2 D-6} \sin^4 \theta} \sin^2 \theta d \widetilde \phi^2  + \cot^2 \theta d\Omega^{D-4} \right)\right\}. \nonumber
\eea
In writing this metric, we have dropped the terms of order 
$\widetilde \rho^2 d \tau d \widetilde \phi$ and various other higher 
order terms in the expansion in $\widetilde \rho$. 
Away from the singular subspace on which $\sin\theta$ vanishes, 
the $(\widetilde \rho, \tau)$ part of the metric is smooth at $\wt\rho=0$
provided  $\tau$ has periodicity 
\be\label{etemp}
(\tau, \widetilde \phi)  \equiv (\tau + \beta, \widetilde \phi) \quad \mbox{where} \quad \beta = 2 \pi \frac{m_0}{b^{D-4}}\, .
\ee 
Using \refb{e443}, this translates to
\be
(\tau, \phi)  \equiv \left(\tau + \beta, \phi-{b^{D-4}\over m_0}\beta\right)\, .
\ee
Of course $\phi$ itself has a periodic identification under shift by $2\pi$ in order that
the metric \refb{ebasemetric} is non-singular for $\rho>b$. From this we can determine the
angular velocity of Euclidean black hole to be
\be\label{e448}
\Omega_E = \frac{b^{D-4}}{m_0}. 
\ee
\refb{etemp} and \refb{e448} give, 
\be
\beta \Omega_E = 2 \pi,
\ee
in agreement with \refb{ebetaomega}.

\subsection{$(D-1)$-dimensional base and the singular subspace}
Remarkably, the string frame metric \refb{emetric} takes the following form in general dimensions\footnote{Ref.~\cite{Horowitz:1995tm} 
also noted this with $\alpha = 0$.}
\be \label{efullmetric}
ds^2  = f (d\tau + \omega_\phi d\phi)^2 + ds^2_{\mathrm{base}},
\ee
where
\be\label{ebasemetric}
ds^2_{\mathrm{base}} = \left(\frac{\rho^2 - b^2 \cos ^2 \theta}{\rho^2 - b^2} \right) d\rho^2 + (\rho^2 - b^2 \cos^2 \theta ) d \theta^2 + (\rho^2 - b^2) \sin^2 \theta d \phi^2 + \rho^2 \cos^2 \theta d \Omega^{D-4},
\ee
\be \label{edeff}
f = \frac{(\rho^2 - b^2 \cos^2 \theta)^2}{\Delta}
= \left(1 + \frac{1}{\rho^{D-5}} \frac{m_0 e^{\alpha}}{\rho^2 - b^2 \cos^2 \theta} \right)^{-1} \left(1 + \frac{1}{\rho^{D-5}} \frac{m_0 e^{-\alpha}}{\rho^2 - b^2 \cos^2 \theta} \right)^{-1},
\ee
and
\be
\omega_\phi = b m_0 \frac{ \rho^{5-D} }{\rho^2 - b^2 \cos^2 \theta} \sin^2 \theta.
\ee
The  base metric $ds^2_{\mathrm{base}}$ does not depend on the charges and
is in fact flat. To see
this, let us introduce,
\bea \label{eydef}
y_1 &=& \sqrt{\rho^2 - b^2} \, \sin \theta \, \cos \phi, \nonumber  \\
y_2 &=& \sqrt{\rho^2 - b^2} \, \sin \theta \,  \sin \phi, \nonumber \\
y_3 &=& \rho\, \cos \theta \cos \psi_1 \nonumber  \\
y_4 & = & \rho \cos\theta \sin \psi_1 \cos \psi_2 \, , \\
& \cdot & \nonumber \\
& \cdot & \nonumber \\
y_{D-2} & = & \rho \cos\theta \sin \psi_1 \cdots \sin \psi_{D-5} \cos
\psi_{D-4} \, , \nonumber \\
y_{D-1} & = & \rho \cos\theta \sin \psi_1 \cdots \sin \psi_{D-5} \sin
\psi_{D-4} \, ,\nonumber
\eea
 where $\psi_1, \ldots \psi_{D-4}$ are the angles labelling points on
the $(D-4)$ sphere.
In this coordinate system the base metric takes the form: 
\be
ds_{\rm base}^2 = dy_1^2 + dy_2^2 + dy_3^2 + \ldots + dy^2_{D-1}.
\ee

Despite having a flat base, the metric is singular is at $\rho=b$, $\cos^2 \theta = 1$
since $f$ vanishes and
$\omega_\phi$ blows up there. To determine the geometry of the singular surface,
note from \refb{eydef} that $y_1$ and $y_2$ vanish on the singular subspace and
$y_3,\cdots, y_{D-1}$ lie on the surface of a $(D-4)$ dimensional sphere of radius $b$.
The range of the coordinate $\theta$ is $(0, \pi)$ for $D=4$ whereas for $D> 4$ it
is $(0, \frac{\pi}{2})$.  Therefore for $D>4$, we have a connected singular surface at
$\rho=b$, $\theta=0$ which is a $S^{D-4}$ sphere. 
In $D=4$ the singular subspace is a pair of points corresponding to $\rho=b$, $\theta=0,\pi$.
Since $S^0$ consists of a pair of points, we can describe the singular surface as $S^{D-4}$
for all $D$. This will be important later.

$f$ computed from \refb{edeff} can be written as
\be
f^{-1} = \left(1 + \frac{m_0 e^\alpha}{2\, \rho^{D-4}} \left(\frac{1}{R_+} + \frac{1}{R_-} \right)\right) \left(1 + \frac{m_0 e^{-\alpha}}{2\, \rho^{D-4}} 
\left(\frac{1}{R_+} + \frac{1}{R_-} \right)\right),
\ee
where
\be
R_\pm = \rho\pm b\cos\theta\, .
\ee
In $D=4$, $R_\pm$ have the interpretation of the distance of a point from the locations
of the singularities at $\rho=b$, $\theta=0,\pi$. In this case 
$f^{-1}$ is simply a product of two harmonic functions and the solution can be written in
a standard IWP form\cite{Boruch:2023gfn,Hegde:2023jmp}.

\sectiono{The geometry near the singularity} \label{ssingular}

Eq.~\refb{e2new} shows that in the two derivative approximation to the effective action,
the macroscopic index vanishes.
According to the 
scaling argument given between \refb{e2} and \refb{e3}, 
higher derivative corrections to $S_{wald}$ from the
non-singular regions cannot produce the answer known from the microscopic counting.
We shall now focus on the singular regions near the horizon and explore whether 
higher derivative correction to $S_{wald}$ from the singular region can produce the
desired result.

\subsection{Coordinates centered at the singular surface}

In order to analyse the solution 
near the singular surface $\cos \theta = 1$, we introduce the following coordinates near the singular subspace. Define,
\be\label{e547}
\widetilde R = \rho  - b \cos \theta,
\qquad  \cos \widetilde \theta  = { \rho \cos \theta - b\over \rho  - b \cos \theta} .
\ee
The inverse coordinate transformations are,
\be
\rho = \frac{1}{2} \left( \widetilde R +  \sqrt{ \widetilde R^2 + 4  b  \widetilde R  \cos \widetilde \theta + 4b^2} \right), \qquad
 \cos \theta  = \frac{1}{2b} \left( - \widetilde R +  \sqrt{  \widetilde R^2 + 4  b \widetilde R \cos \widetilde \theta + 4b^2}\right) .
\ee
$\wt\theta$ has range $(0,\pi)$ for all $D$.
These coordinates have the important property that when $\widetilde R =0$, we have $\rho = b$ \textit{and} $\cos \theta = 1$, i.e., we are at the singular surface. 
For $D=4$ the second copy of the singular subspace is at $\wt R=2b$, $\wt\theta=\pi$,
but near that region it is more appropriate to introduce new coordinate system by
replacing $\cos\theta$ by $-\cos\theta$ in \refb{e547}.


\subsection{Large charge limit}

We now make a final coordinate transformation,
\be
 \widetilde \theta = 2 \Theta, \qquad \widetilde \phi  =  \phi + \frac{\tau}{m_0 b^{4-D}}, 
 \qquad \widetilde \tau = \frac{\tau}{m_0 b^{4-D}},
\ee
and take the limit
$m_0 \to \infty$  and $b \to \infty$  by taking
\be
m_0\sim \lambda^{D-3}, \qquad b\sim \lambda,  \qquad \lambda\to\infty\, ,
\ee 
keeping  fixed $\widetilde R, \widetilde \theta, \wt\tau, \wt\phi,\Omega^{D-4}$.
The scaling of $b$ follows from \refb{e22} after noting that $J_E \propto m_0 b$.
The limit gives
\bea
\rho &\simeq& b + \frac{\widetilde R}{2} (1+\cos\wt\theta)\\
\cos \theta &\simeq& 1- \frac{\widetilde R}{2b} (1 - \cos \widetilde \theta) \\
\sin^2 \theta &\simeq& \frac{\widetilde R}{b} (1 - \cos \widetilde \theta) \\
(\rho^2 - b^2 \cos^2 \theta) &\simeq& 2 b \widetilde R \\
\Delta &\simeq& m_0^2 b^{10-2D} - b^{6-2D}m_0 \widetilde R \left\{
(D-5) b^3 m_0(1+\cos\wt\theta)  - 4 b^D \cosh \alpha \right\} \\
f &\simeq& \frac{4 \widetilde R^2}{m_0^2 b^{8-2D}}
 \\
\omega_\phi &\simeq& m_0 b^{4-D} \sin^2 \frac{\widetilde \theta}{2}.
\eea
We have kept terms in the expansion to the order that will be needed in our computation
of the solution up to terms that vanish for large $\lambda$.
Combining all this together and using  \refb{efullmetric}, we arrive at a universal form
for the metric:
\be
ds^2  \simeq d \widetilde R^2 + 4 \widetilde R^2 d \Theta^2 + 4 \widetilde R^2 \sin^2 \Theta d \widetilde \phi^2 + 4 \widetilde R^2 \cos^2 \Theta d \widetilde \tau^2 + b^2 d \Omega^{D-4}.
\ee
$\Theta$ has range $\left(0, \frac{\pi}{2}\right)$ and $\widetilde \phi$ 
and $\widetilde \tau$ have periodicities $2 \pi$.  The metric is singular at $\widetilde R=0$.
Since $b$ is large, we can
replace $b^2 d \Omega^{D-4}$ with $\sum_{i=1}^{D-4}d x_i^2$ locally.

From this we see that near the singularity 
the metric has a universal form independent of
$D$.
If on the other hand 
we take a spherically symmetric solution by taking
the rotation parameter $b$ to zero, the size of the $(D-4)$ dimensional
sphere goes to zero
and the nature of the singularity becomes dependent on $D$\cite{Peet:1995pe}.

We now turn to the matter fields in the large charge limit at fixed
$\wt R$, $\wt\tau$, $\wt\phi$, $\Theta$. In this limit
\refb{ePQ} gives
\bea 
P & \simeq&   2 \sinh^2 \alpha\, , 
\nonumber \\
U &\simeq& \ -  2  \sinh  \alpha \cosh \alpha\, .
\eea
\refb{eM} now implies,
\be \label{matrix_Mnew}
M \simeq I_{36-2D} +
\pmatrix{
 2 \sinh^2 \alpha \, \,  (n n^T) & -  2  \sinh  \alpha \cosh \alpha
\, \, (n p^T) \cr  -  2  \sinh  \alpha \cosh \alpha  \, \,
 (p n^T)  & 2 \sinh^2 \alpha \, \, (pp^T)}\, .
\ee

For the gauge fields, we get,
for $1\le j\le 26-D$,
\bea 
A_{(E)}^{(j)}{}_{\widetilde \tau} &\simeq&- \sqrt{2} n^{(j)} \cos^2  \Theta  \sinh \alpha \, 
\widetilde R , \nonumber \\
A_{(E)}^{(j)}{}_{\widetilde \phi} &\simeq&- \sqrt{2} n^{(j)} \sin^2  \Theta  \sinh \alpha \, \widetilde R\, ,
\eea
and for $j\ge 27-D$,
\bea 
A_{(E)}^{(j)}{}_{\widetilde \tau} &\simeq& -{p^{(j-26+D)}\over \sqrt 2} b^{4-D} m_0 + \sqrt{2} p^{(j-26+D)} \cos^2  \Theta  \cosh \alpha \, \widetilde R,  \nonumber \\
A_{(E)}^{(j)}{}_{\widetilde \phi} &\simeq& \sqrt{2} p^{(j-26+D)} \sin^2  \Theta  \cosh \alpha \, \widetilde R
\, .
\eea
The constant additive term in the expression for $A_{(E)}^{(j)}{}_{\widetilde \tau}$ 
contributes to the three form field strength
defined in \refb{eH},
but can be
dropped after computing this field strength.

For the two form field, we shall directly write down the expression for the 
gauge invariant three form
field strength after taking into account the contribution from the Chern-Simons term
in \refb{eH}. This takes the form:
\be
H_{(E)} \equiv {1\over 6} H_{(E)\mu\nu\rho} dx^\mu\wedge dx^\nu \wedge dx^\rho
\simeq 8\, \wt R^2 \, \sin\Theta\, \cos\Theta \, d\Theta\wedge d\wt\phi\wedge d\wt\tau\, ,
\ee
where
\be \label{eHEuc}
H_{(E)\mu\nu\rho} \equiv
\p_\mu B_{(E)\nu\rho} + 2 A_{(E)\mu}^{(j)} L_{jk} F^{(k)}_{(E)\nu\rho}
+\p_\nu B_{(E)\rho\mu} + 2 A_{(E)\nu}^{(j)} L_{jk} F^{(k)}_{(E)\rho\mu}
+\p_\rho B_{(E)\mu\nu} + 2 A_{(E)\rho}^{(j)} L_{jk} F^{(k)}_{(E)\mu\nu}\, .
\ee

Finally, the
dilaton behaves as follows,
\be
\exp [- \Phi] \simeq g_s^{-2} \, \frac{b^{4-D} m_0}{2 \widetilde R}.
\ee

If we 
regard this near horizon geometry 
as a solution in four dimensional heterotic string theory
by regarding the coordinates along $\Omega^{D-4}$ as describing an almost flat
compact  space, then the canonical metric in four dimensions is given by $e^{-\Phi}
G_{\mu\nu}$. This describes a flat four dimensional metric, as can be seen by making
a change of coordinates $\wt R=\wt r^2$. The significance of this observation is not clear
to us.

Note that the matrix valued scalar field $M$ and the gauge fields  depend of
the parameter $\alpha$. 
The $\alpha$ dependence on the background $M$ can be removed by doing a 
$O(26-D, 10-D)$ transformation described in \refb{eodd} by choosing,
\be\label{eorot}
\Omega = I_{36-2D}+\pmatrix{
 (\cosh \alpha-1) \, \,  (n n^T) &   \sinh  \alpha 
\, \, (n p^T) \cr     \sinh  \alpha  \, \,
 (p n^T)  & (\cosh \alpha-1) \, \, (pp^T)
}\, .
\ee
This gives 
\be
M \simeq I_{36-2D}.
\ee
The vector fields also change under this.  We get, for $1\le j\le 26-D$,
\bea 
A{}{}_{(E)}^{(j)}{}_{\widetilde \tau} &\simeq&0, \nonumber \\
A{}{}_{(E)}^{(j)}{}_{\widetilde \phi }{}&\simeq& 0\, ,
\eea
and for $j\ge 27-D$,
\bea 
A{}{}_{(E)}^{(j)}{}_{\widetilde \tau} &\simeq&\sqrt{2} p^{(j-26+D)}  \widetilde R\cos^2  \Theta,
\nonumber \\
A{}{}_{(E)}^{(j)}{}_{\widetilde \phi} &\simeq& \sqrt{2} p^{(j-26+D)}  \widetilde R\sin^2  \Theta\, ,
\eea
where we have dropped the constant term in the expression for
$A{}{}_{(E)}^{(j)}{}_{\widetilde \tau}$. Since we no longer need to use the gauge field
for computing $H_{(E)}$, this is an allowed operation.
The new fields do not depend on the parameter $\alpha$. 
Furthermore using an $O(10-D)\subset O(26-D,10-D)$ rotation we could bring the unit
vector $p$ to any standard form, {\it e.g.} having only the first component non-zero.

\sectiono{Wald entropy and the index} \label{swald}

Let us summarize the solution in the large charge limit after the $O(26-D,10-D)$ rotation
\refb{eorot}. We have,
\ben\label{eF1}
ds^2  &\simeq& d \widetilde R^2 + 4 \widetilde R^2 d \Theta^2 + 4 \widetilde R^2 \sin^2 \Theta d \widetilde \phi^2 + 4 \widetilde R^2 \cos^2 \Theta d \widetilde \tau^2 + b^2 d \Omega^{D-4},
\nonumber \\
&& \hskip 1in 0\le\Theta\le {\pi\over 2}, \qquad \wt\phi\equiv \wt\phi+2\pi, \qquad 
\wt\tau\equiv \wt\tau+2\pi\, ,
\een
\be\label{eF2}
M \simeq I_{36-2D}\, ,
\ee
\bea \label{eF3}
A{}{}_{(E)}^{(j)}{}_{\widetilde \tau} &\simeq&0, \nonumber \\
A{}{}_{(E)}^{(j)}{}_{\widetilde \phi }{}&\simeq& 0\, , \qquad \hbox{for $1\le j\le 26-D$}\, ,
\eea
\bea \label{eF4}
A{}{}_{(E)}^{(j)}{}_{\widetilde \tau} &\simeq&\sqrt{2} p^{(j-26+D)}  \widetilde R\cos^2  \Theta,
\nonumber \\
A{}{}_{(E)}^{(j)}{}_{\widetilde \phi} &\simeq& \sqrt{2} p^{(j-26+D)}  \widetilde R\sin^2  \Theta\, ,\qquad \hbox{for $j\ge 27-D$}\, ,
\eea
\be\label{eF5}
H_{(E)}\simeq
8\, \wt R^2 \, \sin\Theta\, \cos\Theta \, d\Theta\wedge d\wt\phi\wedge d\wt\tau\, ,
\ee
and 
\be\label{eF6}
\exp [- \Phi] \simeq g_s^{-2}\, \frac{b^{4-D} m_0}{2 \widetilde R}\, .
\ee
We now note the following features of the solution:
\begin{enumerate}
\item Since the original configuration described in section \ref{sbps} solved the equations
of motion of the two derivative theory for all values of the parameters, and the
configuration given in \refb{eF1}-\refb{eF6} is obtained as a limit of the original solution
for large $m_0$, $b$, \refb{eF1}-\refb{eF6} should also satisfy the equations of
motion of the two derivative theory. We have checked this by explicit computation, after
replacing the $b^2d\Omega^{D-4}$ part of the metric by flat metric.
However since for 
finite $\wt R$ the curvature invariants and other invariants 
associated with this solution are of order unity, we expect the solution to be modified by
higher derivative corrections. Such corrections should fall off for large $\wt R$
where the invariants are small.

\item In the large charge limit, since $b\sim\lambda$ and $m_0\sim \lambda^{D-3}$,
$e^{-\Phi}$ given in \refb{eF6} is of order $\lambda$. Since $e^{-\Phi}$
can be identified as the inverse square of the string coupling locally, 
we see that near the singularity the string coupling is small.
Hence we can ignore string loop corrections and focus only on the $\alpha'$ correction to
the solution.
\item  Since $\alpha'$ correction to the effective action does not destroy the $O(26-D, 10-D)$
symmetry, the $O(26-D, 10-D)$ transformation that was used to bring the solution to the
form \refb{eF1}-\refb{eF6} is a valid symmetry and leaves the $\alpha'$ corrected formula for
$S_{wald}$ invariant. In fact we can use the $O(10-D)$ subgroup of $O(26-D, 10-D)$ to
bring the vector $p$ to some canonical form, {\it e.g.} only with the first component non-zero.
In this case the solution is universal except for the $b^2 d\Omega^{D-4}$ term in the metric 
\refb{eF1} and
the overall factor of $g_s^{-2}\,
b^{4-D}m_0$ in the expression for $e^{-\Phi}$ in \refb{eF6}. In the
limit of large $b$ the $b^2 d\Omega^{D-4}$ part of the metric describes almost locally flat
$(D-4)$ dimensional space and the $\alpha'$ corrections are suppressed by inverse power of
$\lambda$. At the leading order this part of the metric will give an overall multiplicative factor
of $b^{D-4}\omega_{D-4}$ to the action and hence to $S_{wald}$ but has no effect on the
$\alpha'$ correction to the rest of the field configuration. On the other hand from
\refb{action} we see that the 
effect of the  $g_s^{-2} \, b^{4-D}m_0$ in the expression for $e^{-\Phi}$ is to multiply the action by
a factor of $g_s^{-2}\, 
b^{4-D}m_0$ but it also does not affect the $\alpha'$ corrections to the solution since these
corrections involve $\Phi$ only through its derivatives.  
Therefore $\alpha'$ corrections to the solution take a universal form, even independent
of $D$, and modify only the $\widetilde R$ and $\Theta$ dependence of various fields in
\refb{eF1}-\refb{eF6}.
\end{enumerate}

Taking into account the overall factor of $C_D$ in \refb{action} we
see that the net $\alpha'$ correction to $S_{wald}$, and hence to $S_{macro}$ according to
\refb{efull}, \refb{e2}, takes the form,
\be\label{esmacfin}
S_{macro}=K\times  C_D\times b^{D-4}\omega_{D-4} \times g_s^{-2}\, 
b^{4-D}m_0 = K\, C_D \, g_s^{-2}\, m_0 \omega_{D-4} \, ,
\ee
where $K$ is a universal numerical constant that can in principle be computed from $\alpha'$
correction to the solution \refb{eF1}-\refb{eF6}. In particular, $K$ is independent of $D$.

We now recall from \refb{echarges}:
\be\label{echargesnew}
\sqrt{Q_R^2 - Q_L^2} = \sqrt{Q^T L Q} = \sqrt 2 \, (D-3)\, C_D \, g_s^{-2}\, 
m_0\, \omega_{D-2} \, .
\ee
Using this we can express \refb{esmacfin} as
\be\label{esmac1}
S_{macro}={K\over \sqrt 2} {\omega_{D-4}\over (D-3)\omega_{D-2}} \,
\sqrt{Q_R^2 - Q_L^2}  \, .
\ee
Using the result
\be
\omega_k = 2 \pi^{(k+1)/2} / \Gamma[(k+1)/2]\, ,
\ee
we get
\be 
\omega_{D-2}/\omega_{D-4} = 2\pi / (D-3)\, .
\ee
Substituting this into \refb{esmac1} we get
\be\label{esmac2}
S_{macro} = {K\over 2\pi \sqrt 2}\, \sqrt{Q_R^2 - Q_L^2}\, .
\ee
This would agree with \refb{emicro} if $K=8\pi^2$. While we do not have a
way to calculate $K$ at present, the $D$-independence of \refb{esmac2} is
consistent with the $D$-independence of \refb{emicro}.

Note that the dependence on the parameter $b$, that controls the temperature of
the black hole via \refb{etemp}, has disappeared. This is consistent with the fact that
the index should be independent of the temperature. Note also that the result is
independent of the combination $Q^TM_\infty Q=Q_L^2+Q_R^2$. This can be traced
to the $\alpha$-independence of the result, -- the $O(26-D,10-D)$ transformation 
that removes the $\alpha$ dependence of $M$ near the singularity
also removes the $\alpha$ dependence
of the gauge fields. Such a dependence would have been inconsistent with the
microscopic results which do not depend on $M_\infty$. This was also a feature in the
analysis in \cite{Sen:1995in}.

\sectiono{Generalization to other compactifications} \label{sgen}

The above analysis can be easily generalized to other $\NN=4$ or $\NN=2$
supersymmetric 
compactifications of heterotic string theory. For this we use the observation of
\cite{Sen:1997is} that the black hole solution describing BPS
elementary string states can be embedded  in a universal consistent truncation of the
supergravity theory used above. Furthermore the universality is maintained 
to all orders in $\alpha'$. Therefore the analysis given above can be carried
out without any further modification and we shall arrive at the expression
\refb{esmac2} for the macroscopic entropy with the same $K$. The microscopic
counting also remains unchanged, since the asymptotic growth of BPS states
is controlled by the central charge of the left-moving matter sector of the world-sheet
theory
and this is always equal to 26. Note however that in order to get BPS fundamental string
in the first place, we must have at least one internal circle on which the string can
wrap, i.e.\ we cannot have such states in ten dimensional heterotic string theory or
the six dimensional theory obtained by compactifying heterotic string theory on
$K3$.

Finally we shall make a few remarks on the type II string compactification.
For definiteness let us consider the case of torus compactification. 
The microscopic
index for the fundamental string states
vanishes in this case due to the cancellation between the contributions from the
bosonic and fermionic states.
On the macroscopic side the relevant part of the two derivative action is very similar to
the one we have analyzed here and identical arguments lead to a form \refb{esmac2}
for the logarithm of the index. Therefore for consistency the coefficient $K$ must vanish for
type II theory. This may be related to the fact that higher derivative corrections are relatively
rare in type II string theory. For example the four derivative term, that was used in 
\cite{Dabholkar:2004yr} to get a finite result for $K$ in the heterotic string theory,
is absent in type II
string theory. However to get a complete picture we need to learn how to compute $K$
by taking into account the effect of all the $\alpha'$ corrections to the effective action.

\bigskip

\noindent{\bf Acknowledgement}: 
We would like to thank A H Anupam for discussions and collaboration during the
initial stages of this work.
CC thanks David Turton and Shasha Tyukov for discussions.
AS thanks Sameer Murthy for discussions.
AV thanks David Chow for email correspondence.
PS would like to thank the Raman Research Institute and the International Centre for Theoretical Sciences for their kind hospitality during the initial stages of this work.
CC is supported by the STFC consolidated grant (ST/X000583/1) ``New Frontiers in Particle Physics, Cosmology and Gravity''.
A.S. is supported by ICTS-Infosys Madhava 
Chair Professorship
and the J.~C.~Bose fellowship of the Department of Science and Technology.




\begin{thebibliography}{99}


\bibitem{tHooft:1990fkf}
G.~'t Hooft,
``The black hole interpretation of string theory,''
Nucl. Phys. B \textbf{335} (1990), 138-154
doi:10.1016/0550-3213(90)90174-C

\bibitem{Susskind:1993ws}
L.~Susskind,
``Some speculations about black hole entropy in string theory,''
[arXiv:hep-th/9309145 [hep-th]].

\bibitem{Susskind:1994sm}
L.~Susskind and J.~Uglum,
``Black hole entropy in canonical quantum gravity and superstring theory,''
Phys. Rev. D \textbf{50} (1994), 2700-2711
doi:10.1103/PhysRevD.50.2700
[arXiv:hep-th/9401070 [hep-th]].

\bibitem{Russo:1994ev}
J.~G.~Russo and L.~Susskind,
``Asymptotic level density in heterotic string theory and rotating black holes,''
Nucl. Phys. B \textbf{437} (1995), 611-626
doi:10.1016/0550-3213(94)00532-J
[arXiv:hep-th/9405117 [hep-th]].

\bibitem{Horowitz:1996nw}
G.~T.~Horowitz and J.~Polchinski,
``A Correspondence principle for black holes and strings,''
Phys. Rev. D \textbf{55} (1997), 6189-6197
doi:10.1103/PhysRevD.55.6189
[arXiv:hep-th/9612146 [hep-th]].

\bibitem{Chen:2021dsw}
Y.~Chen, J.~Maldacena and E.~Witten,
``On the black hole/string transition,''
JHEP \textbf{01} (2023), 103
doi:10.1007/JHEP01(2023)103
[arXiv:2109.08563 [hep-th]].


\bibitem{Dabholkar:1989jt}
A.~Dabholkar and J.~A.~Harvey,
``Nonrenormalization of the Superstring Tension,''
Phys. Rev. Lett. \textbf{63} (1989), 478
doi:10.1103/PhysRevLett.63.478


\bibitem{Sen:1995in}
A.~Sen,
``Extremal black holes and elementary string states,''
Mod. Phys. Lett. A \textbf{10} (1995), 2081-2094
doi:10.1142/S0217732395002234
[arXiv:hep-th/9504147 [hep-th]].

\bibitem{Peet:1995pe}
A.~W.~Peet,
``Entropy and supersymmetry of D-dimensional extremal electric 
black holes versus string states,''
Nucl. Phys. B \textbf{456} (1995), 732-752
doi:10.1016/0550-3213(95)00537-2
[arXiv:hep-th/9506200 [hep-th]].

\bibitem{Sen:1997is}
A.~Sen,
``Black holes and elementary string states in N=2 supersymmetric string theories,''
JHEP \textbf{02} (1998), 011
doi:10.1088/1126-6708/1998/02/011
[arXiv:hep-th/9712150 [hep-th]].

\bibitem{Dabholkar:2004yr}
A.~Dabholkar,
``Exact counting of black hole microstates,''
Phys. Rev. Lett. \textbf{94} (2005), 241301
doi:10.1103/PhysRevLett.94.241301
[arXiv:hep-th/0409148 [hep-th]].

\bibitem{Sen:2005ch}
A.~Sen,
``Black holes and the spectrum of half-BPS states in N=4 supersymmetric string theory,''
Adv. Theor. Math. Phys. \textbf{9} (2005) no.4, 527-558
doi:10.4310/ATMP.2005.v9.n4.a1
[arXiv:hep-th/0504005 [hep-th]].


\bibitem{2107.09062}
L.~V.~Iliesiu, M.~Kologlu and G.~J.~Turiaci,
``Supersymmetric indices factorize,''
JHEP \textbf{05} (2023), 032
doi:10.1007/JHEP05(2023)032
[arXiv:2107.09062 [hep-th]].

\bibitem{Cabo-Bizet:2018ehj}
A.~Cabo-Bizet, D.~Cassani, D.~Martelli and S.~Murthy,
``Microscopic origin of the Bekenstein-Hawking entropy of supersymmetric AdS$_{5}$ black holes,''
JHEP \textbf{10} (2019), 062
doi:10.1007/JHEP10(2019)062
[arXiv:1810.11442 [hep-th]].

\bibitem{Boruch:2023gfn}
J.~Boruch, L.~V.~Iliesiu, S.~Murthy and G.~J.~Turiaci,
``New forms of attraction: Attractor saddles for the black hole index,''
[arXiv:2310.07763 [hep-th]].


\bibitem{H:2023qko}
A.~H.~Anupam., P.~V.~Athira, C.~Chowdhury and A.~Sen,
``Logarithmic Correction to BPS Black Hole Entropy from Supersymmetric Index at Finite Temperature,''
[arXiv:2306.07322 [hep-th]].

\bibitem{Anupam:2023yns}
A.~H.~Anupam, C.~Chowdhury and A.~Sen,
``Revisiting Logarithmic Correction to Five Dimensional BPS Black Hole Entropy,''
[arXiv:2308.00038 [hep-th]].

\bibitem{Wald:1993nt}
R.~M.~Wald,
``Black hole entropy is the Noether charge,''
Phys. Rev. D \textbf{48} (1993) no.8, R3427-R3431
doi:10.1103/PhysRevD.48.R3427
[arXiv:gr-qc/9307038 [gr-qc]].



\bibitem{Maharana:1992my}
J.~Maharana and J.~H.~Schwarz,
``Noncompact symmetries in string theory,''
Nucl. Phys. B \textbf{390}, 3-32 (1993)
doi:10.1016/0550-3213(93)90387-5
[arXiv:hep-th/9207016 [hep-th]].

\bibitem{Sen:1991zi}
A.~Sen,
``O(d) x O(d) symmetry of the space of cosmological 
solutions in string theory, scale factor duality and two-dimensional black holes,''
Phys. Lett. B \textbf{271} (1991), 295-300
doi:10.1016/0370-2693(91)90090-D

\bibitem{Hassan:1991mq}
S.~F.~Hassan and A.~Sen,
``Twisting classical solutions in heterotic string theory,''
Nucl. Phys. B \textbf{375} (1992), 103-118
doi:10.1016/0550-3213(92)90336-A
[arXiv:hep-th/9109038 [hep-th]].


\bibitem{Horowitz:1995tm}
G.~T.~Horowitz and A.~Sen,
``Rotating black holes which saturate a Bogomolny bound,''
Phys. Rev. D \textbf{53}, 808-815 (1996)
doi:10.1103/PhysRevD.53.808
[arXiv:hep-th/9509108 [hep-th]].

\bibitem{Emparan:2008eg}
R.~Emparan and H.~S.~Reall,
``Black Holes in Higher Dimensions,''
Living Rev. Rel. \textbf{11} (2008), 6
doi:10.12942/lrr-2008-6
[arXiv:0801.3471 [hep-th]].



\bibitem{Hegde:2023jmp}
S.~Hegde and A.~Virmani,
``Killing Spinors for Finite Temperature Euclidean Solutions at the BPS Bound,''
[arXiv:2311.09427 [hep-th]].



\end{thebibliography}
\end{document}